# Extraction of Material Properties through Multi-fidelity Deep Learning from Molecular Dynamics Simulation

Mahmudul Islam[1], Md Shajedul Hoque Thakur[1], Satyajit Mojumder[2] and Mohammad Nasim Hasan[1,*]

[1]Department of Mechanical Engineering, Bangladesh University of Engineering and Technology, Dhaka-1000, Bangladesh.

[2]Theoretical and Applied Mechanics Program, Northwestern University, Evanston, IL-60208, USA

[*]Corresponding author *Email address:* nasim@me.buet.ac.bd

# Abstract

Simulation of reasonable timescales for any long physical process using molecular dynamics (MD) is a major challenge in computational physics. In this study, we have implemented an approach based on multi-fidelity physics informed neural network (MPINN) to achieve long-range MD simulation results over a large sample space with significantly less computational cost. The fidelity of our present multi-fidelity study is based on the integration timestep size of MD simulations. While MD simulations with larger timestep produce results with lower level of accuracy, it can provide enough computationally cheap training data for the MPINN to learn an accurate relationship between these low-fidelity results and high-fidelity MD results obtained using smaller simulation timestep. We have performed two benchmark studies, involving one and two component Lennard-Jones systems, to determine the optimum percentage of high-fidelity training data required to achieve accurate results with high computational saving. The results show that important system properties such as system energy per atom, system pressure and diffusion coefficients can be determined with high accuracy while saving 68% computational costs. Finally, as a demonstration of the applicability of our present methodology in practical MD studies, we have studied the viscosity of argon-copper nanofluid and its variation with temperature and volume fraction by MD simulation using MPINN. Then we have compared them with numerous previous studies and theoretical models. Our results indicate that MPINN can predict accurate nanofluid viscosity at a wide range of sample space with significantly small number of MD simulations. Our present methodology is the first implementation of MPINN in conjunction with MD simulation for predicting nanoscale properties. This can pave pathways to investigate more complex engineering problems that demand long-range MD simulations.

*Keywords:* Molecular dynamics, Multi-fidelity models, Deep Neural Networks, Nanofluid, Material Properties.

# 1. Introduction

Molecular dynamics (MD) is a computational method that simulates the time evolution of a set of interacting atoms or molecules using Newton's equations of motion. MD simulation has displayed outstanding predictive power in simulating a wide range of nanoscale phenomena[1,2]. With the significant progress in nanotechnology and improved computational capability in recent years, MD simulation has become one of the most widely used computational tools among scientists and researchers. Despite of its excellent predictive capability, there are several design constraints of MD simulation. Design constraints of any simulation method are the maximum or minimum values of a particular simulation parameter beyond which the accuracy and stability of the simulation are significantly reduced. The main bottlenecks of MD simulations are the simulation system size (number of atoms), timestep and total time duration, which must be selected according to available computational resources so that the simulation can be finished within a reasonable time period. To make the simulation statistically valid and conclusive, the total simulation duration should be long enough to replicate the natural processes being studied. It is almost impossible to simulate certain long physical processes accurately within reasonable time period, even with the most powerful supercomputer. This particular issue should be addressed properly, when we are dealing with systems involving rough energy landscape and long relaxation processes. One way to address this issue is by increasing the integration timestep, which allows to simulate longer physical processes within a reasonable time period. However, it comes with penalty on accuracy and stability of the numerical process.

Several approaches have been proposed in the literature to address and solve the design constraints of long-range MD simulation. By applying transition-state-theory[3], one can compute the rate of infrequent events during a physical process and eventually, estimate long range MD results with reasonable accuracy. However, this approach requires prior knowledge of all the important reaction paths, which is very difficult to obtain for complex nanoscale processes. Hence, Hamelberg *et al.*[4] proposed a robust bias potential function that can accelerate MD simulations by simulating the transition of high energy barriers without any prior knowledge of the location of potential energy wells or saddle points. This approach along with other similar approaches[5,6] characterized by interatomic energy manipulation, is known as accelerated molecular dynamics (AMD). However, implementation of AMD requires performing a series of rigorous computational procedures informed by in-depth understanding of the underlying MD processes. In recent years, predictive algorithms have been implemented in the area of MD to achieve accurate and computationally efficient long-range MD results. Reeve *et al.*[7] applied functional uncertainty quantification (UQ) to Lennard-Jones (two-body) interaction model to predict high order interatomic interactions. This approach is computationally expensive since it requires large number of samples from low-fidelity models. Also, this approach is applicable only when the discrepancies between low and high-fidelity potential energy stay within acceptable bounds. Researchers have also focused on optimizing force field parameters in order to conduct MD simulations for complex and far-from-equilibrium chemical processes within reasonable time [8–10]. Mishra *et al.*[10] introduced a dynamic approach based on multi-objective genetic algorithm to train the ReaxFF[11] potential parameters, which significantly reduces reactive molecular dynamics (RMD) simulation time. Pilania *et al.*[12] recently proposed a variable fidelity machine learning regression model to calculate bandgaps of solid materials. This model is based upon Gaussian process regression framework which requires sufficient independent data for parameter estimation. Also, the validity of this model is predicated upon the assumption that, Gaussian process can describe the underlying molecular dynamics properly. Razi *et al.*[13] introduced a novel approach based upon the construction of a multi-fidelity surrogate model and conducted a benchmark study on two Lennard Jones (LJ) systems using MD simulation models having different levels of accuracy. In this study, we aim to apply deep neural network (DNN) based on multi-fidelity sampling to get accurate and efficient prediction of MD simulation results.

In recent years, we have witnessed several pioneering advancements in the fields of machine learning (ML) and neural network (NN)[14–16]. Typically, a neural network can be characterized as a series of algorithms that can recognize the underlying relationships between a set of data accurately through a process that mimics the way the human brain operates. The accuracy of the NN depends especially on the accuracy of the available training data. Training data are extracted from physically realistic models of a system or process with different degrees of complexity. The idea of coupling computational physics and NN is fairly new and bears great promise. It is apparent from recent studies[12] that, a combination of simulation results of different levels of fidelity can significantly reduce the computational cost of training data generation for NN. Also different previous studies[17–20] leveraged training data extracted from multiple models of different fidelity to achieve higher accuracy and computational efficiency in NN. Lu *et al.*[21] used multi-fidelity neural network to integrate simulation and experimental data of nanoindentation of commercial alloys for training disparate datasets to learn and minimize systematic errors. They used the multi-fidelity physics informed neural network (MPINN) developed by Meng *et al.*[22]. MPINN can also be used to combine simulation results of different fidelity models to predict system properties accurately. In their multi-fidelity framework, a small number of training data from computationally expensive but accurate high-fidelity simulations can be combined with a large number of training data from computationally cheaper but less accurate low-fidelity simulations to greatly enhance the NN prediction accuracy. Also, the method of multi-fidelity sampling using NN does not require any prior knowledge of the physical processes or simulation model. By incorporating MD simulations' integration timestep as the determinant of model fidelity, it is possible to acquire fairly accurate and computationally inexpensive long-range MD results using MPINN.

In our present study, we have implemented the multi-fidelity physics informed neural network (MPINN), to predict system properties using MD results of two different fidelity, where the fidelity is based on the size of integration timestep (section 2.1). We have considered highly accurate MD simulation results with small timestep as high-fidelity data and relatively inaccurate but computationally cheap results obtained from MD simulations with large integration timestep as low-fidelity data. The MPINN has been implemented to predict different properties of two canonical Lennard-Jones (LJ) systems (section 3), as benchmark studies. The accuracy (section 3.2.1 and 3.2.2) and computational saving (section 3.2.3) of our present method have been extensively discussed. Finally, we applied the MPINN to predict the viscosity of argon-copper nanofluid for different temperatures and volume fractions (section 4). We compared the MPINN predictions with available literature (section 4.2) in order to present the excellent computational efficiency and accuracy that can be achieved using our present methodology.

## 2. Multi-fidelity physics informed neural network

Defining the basis of fidelity is the key starting point for any multi-fidelity study. In MD studies, when we know the state configuration at a particular moment, the state of the next moment can be obtained using discrete integration method on the Newtonian equations of motion. Several algorithms have been developed to conduct this integration process [23–25]. In all these algorithms, the most important parameter is the simulation timestep. Computationally, it is most efficient to use the largest timestep possible for simulating the long processes. Small timestep sizes result in long computation time which is not desirable for simulating long physical or chemical processes. However, the algorithm becomes unstable when large timestep is used. In this case, the motion of particles become unstable due to large truncation error in the integration process which results in rapid increase in total energy. The system may lead to unphysical structures due to this. This behavior is called exploding and is caused by devastating atomic collisions that occurs when large timestep propagates the positions of two atoms to be nearly overlapping creating a strong repulsive force. Even if the simulation is not unstable, large timestep MD simulations are still very inaccurate compared to small timestep MD simulations. So, large timestep is computationally efficient but gives inaccurate results, whereas small timestep is highly accurate but computationally expensive. In our present multi-fidelity implementation, we have considered results (various system properties) obtained from large timestep MD simulations as low-fidelity data and results obtained from small timestep MD simulations as high-fidelity data.

In any multi-fidelity model, the low-fidelity data are expected to have errors associated with them, but none of the low-fidelity results can be divergent. Figure 1 illustrates, total energy per atom of a 1000 atoms LJ system during equilibration obtained using MD simulation for different timesteps. As we can see from the figure a significant computational saving can be obtained by increasing the simulation timestep. However, it is also clear from the figure that, smaller timestep (1 fs and 10 fs) MD simulations yield a more stable system energy per atom compared to larger timesteps. For 200 fs timestep, the system energy per atom fluctuates a lot, but it still has a stable mean, which indicates that this particular timestep is highly erroneous but not divergent. With the increase in timestep, fluctuations of the system energy per atom increases. But as long as it is not divergent (increasing or decreasing rapidly) we can use that timestep to generate meaningful low fidelity results. All the low-fidelity MD simulation timesteps in our present study have been chosen ensuring that the simulation does not become divergent.

For any multi-fidelity modelling, one of the main objectives is to determine the relationship between low- and high-fidelity data.[26] In some previous studies,[26] this relationship has been expressed as:

$$y_H = \rho(x)y_L + \delta(x) \qquad (1)$$

where, $y_H$ and $y_L$ are high- and low-fidelity data respectively, $\rho(x)$ is the multiplicative correlation factor and $\delta(x)$ is the additive correlation factor. One of the main disadvantages of this relationship is that, it can only handle linear correlation between two fidelity data, whereas a lot of interesting cases[27,28] follow non-linear relationships between low- and high-fidelity data. For those cases, the relationship between low- and high-fidelity data are expressed as:

$$y_H = F(x, y_L) \qquad (2)$$

where, $F(.)$ is an unknown (nonlinear/linear) function that correlates low fidelity data to high-fidelity data. Meng *et al.*[22] explored the linear/nonlinear correlation by decomposing $F(.)$ into linear and nonlinear parts, which are expressed as:

$$F = F_l + F_{nl} \tag{3}$$

where, $F_l$ and $F_{nl}$ are the linear and nonlinear terms in $F$, respectively. Therefore, the correlation between low- and high-fidelity data becomes:

$$y_H = \alpha F_l(x, y_L) + (1 - \alpha)F_{nl}(x, y_L), \alpha \in [0,1] \tag{4}$$

where, $\alpha$ is a hyper-parameter which determines the degree of nonlinearity between low- and high-fidelity data. This hyper-parameter along with the linear and non-linear relationships have to be determined by training the MPINN.

The architecture of the MPINN used in our present study was proposed and validated by Meng *et al.*[22] for a wide range of physical problems. As shown in Fig. 2, the MPINN is composed of three fully connected neural networks. The first one ($NN_L$) approximates the low fidelity data. The second ($NN_{H1}$) and third ($NN_{H2}$) neural networks, which together has been denoted as $NN_H$, approximate the linear ($F_l$) and nonlinear ($F_{nl}$) correlations for the low- and high-fidelity data, respectively. The size (depth and width) of the MPINN has a strong effect on the predictive accuracy. Usually, sufficient number of low-fidelity data is available due to its low computational cost. Hence, it is easy to find an appropriate size for $NN_L$ to approximate the low-fidelity function. But since few computationally expensive high-fidelity data is available, particular focus should be put on the size of $NN_H$ while designing MPINN. Meng *et al.* [22] investigated this and suggested the optimum ranges of the depth ($l$) and width ($w$) of $NN_{H2}$ as: $l \in [1,2]$ and $w \in [4,20]$. $NN_{H2}$ of our present MPINN consists of 2 hidden layers ($l = 2$) with 20 neurons in each hidden layer ($w = 20$). Also, as $NN_{H1}$ only predicts the linear relationship between high- and low-fidelity data, one hidden layer with one neuron in $NN_{H1}$ is sufficient. The specifications of the neural networks are presented in Table 1.

**Table 1.** Specifications of the NNs present in MPINN.

| Neural Network | Number of hidden layers | Neurons per hidden layer | Activation function | $L_2$ regularization parameter |
|---|---|---|---|---|
| $NN_L$ | 4 | 20 | Tangent sigmoid function in input and hidden layers + linear function in the output layer | 0 |
| $NN_{H1}$ | 1 | 1 | Linear function | 0 |
| $NN_{H2}$ | 2 | 20 | Tangent sigmoid function | 0.001 |

The loss function that has been optimized in the present study is:

$$MSE = \frac{1}{N_{yL}} \sum_{i=1}^{N_{yL}} (y_L^* - y_L)^2 + \frac{1}{N_{yH}} \sum_{i=1}^{N_{yH}} (y_H^* - y_H)^2 \qquad (5)$$

where, $y_L^*$ and $y_H^*$ denotes the output of the $NN_L$ and $NN_H$, respectively. To avoid overfitting, we have used the standard $L_2$ regularization[29] only in $NN_{H2}$ neural network since a small number of available high-fidelity data may yield overfitting. The value of regularization parameter used in this neural network is 0.001, which has been found to avoid overfitting across a wide range of dataset and also ensure no significant underfitting. The loss function has been optimized using L-BFGS method[30] along with Xavier's initialization method[31] and the learning rate has been set to 0.001. For our present benchmark study, temperature and density are considered as the input parameters (*x*) of the MPINN and total energy per atom, system pressure and self-diffusion coefficient (details in section 2.2) are the desired output (*y*) that the MPINN is trained to accurately predict. For the nanofluid viscosity study, temperature and volume fraction are considered as the input parameters (*x*) of the MPINN and viscosity (details in section 2.3) is the desired output (*y*). All versions of MPINNs of our present study are implemented in MATLAB[32].

## 3. Benchmark Study

### 3.1 Benchmark molecular dynamics simulation model

To demonstrate the predictive capability of the MPINN proposed in the present study for multi-fidelity MD simulations, we have selected two different benchmark systems. One is a unary system with single component and the other is a binary system consisting of two different components.

### 3.1.1 Single Component System

The first benchmark system is a one-component Lennard–Jones (LJ) system of 1000 atoms. As shown in Fig. 3(a), the simulation domain is a cubic box with periodic boundary conditions in all sides. The size of the simulation domain varies according to the density of the system, since density is a parameter of the study. The interactions between the atoms in the unary system are described by the well-known Lennard–Jones (LJ) 12-6 potential[33]:

$$u(r_{ij}) = 4\epsilon \left[ \left( \frac{\sigma}{r_{ij}} \right)^{12} - \left( \frac{\sigma}{r_{ij}} \right)^{6} \right] \tag{6}$$

where $r_{ij}$ is the distance between atom $i$ and atom $j$, $\sigma$ is the distance at which the inter-particle potential is zero and $\varepsilon$ is the depth of the potential well. For this system, we assume $\sigma$ and $\varepsilon$ to be 3 Å and 0.04336 eV, respectively. The atomic mass of the atoms has been set to 83.798. It is worth mentioning that, the potential parameters and atomic mass in this benchmark system is not of any particular element in the periodic table. We have selected these parameters so that, the system is a fairly well representation of a single element LJ system. If the MPINN can accurately predict different properties of this system, then it is expected to be applicable for a wide variety of single component LJ systems.

### 3.1.2 Two Component System

Figure 3(b) illustrates the second benchmark system, which is a two-component (component A and component B) LJ system of 1000 atoms, each component consisting of 500 atoms. Similar to the single component system, the simulation domain is a cubic box with periodic boundary conditions in all sides and its dimensions depending on the density of the system. The interactions between the atoms in the binary system are described by the well-known Lennard–Jones (LJ) 12-6 potential (equation 6). The values of the interaction potential parameters used for this two-component system are listed in Table 2.

**Table 2.** LJ (12-6) interatomic potential parameters for two-component benchmark system

| Particles ($i, j$) | $\varepsilon_{ij}$ (eV) | $\sigma_{ij}$ (Å) |
|---|---|---|
| A-A | 0.0104 | 3.5 |
| B-B | 0.0104 | 3.89 |
| A-B | 0.0104 | 3.69 |

It is worth mentioning that, the interaction parameters between A and B have been calculated using Lorentz-Berthelot rules[34]. The atomic masses of A and B are 131.293 and 83.798, respectively. This binary system is a good enough representation of two component LJ systems. The aim of studying this system is to determine whether our proposed MPINN is applicable for accurately predicting different system properties of two component systems.

### 3.1.3 Simulation Procedure of Benchmark Study

Initially, for both one-component and two-component systems, the simulation domain has been thermally equilibrated for 100 ps using the canonical (NVT) ensemble at desired temperature. After the initial equilibration, the system has been kept at a desired temperature for 2 ns using canonical (NVT) ensemble to calculate different system properties such as total system energy per atom, system pressure and diffusion coefficient. For both implementation of NVT ensemble, the temperature damping parameter is 10 fs and the drag factor is set to 0.3. The system energy per atom and pressure fluctuate during this 2 ns period. Hence, we have considered the median values of system energy per atom and pressure over 2 ns period as training data for the MPINN. To calculate diffusion coefficient, we first evaluated mean squared displacement (MSD) of the particles using this formula[35]:

$$MSD(t) = \frac{1}{N} \sum_{i=1}^{N} \langle |r_i(t) - r_i(0)|^2 \rangle \tag{7}$$

where, $r_i(0)$ and $r_i(t)$ are the position coordinates of $i^{th}$ atom at the initial time and current time, respectively. The limiting slope of *MSD(t)*, considered for sufficiently long time interval, is related to the self-diffusion coefficient (*D*) by following formula[35]:

$$D = \frac{1}{6} \lim_{t \to \infty} \frac{d(MSD(t))}{dt} \tag{8}$$

Since the MSD vs time curves in our present study are linear, the diffusion coefficient, *D*, can be simply calculated as 1/6$^{th}$ of the slope of the MSD vs time curves.

To evaluate different system properties at different system conditions, we have considered a uniform grid of temperature and density. Temperatures from 500 K to 1000 K at 50 K intervals and densities of 36.27 kg/m$^3$ to 701.29 kg/m$^3$ at 66.5 kg/m$^3$ intervals have been taken as the parameters of this study, for both one-component and two-component benchmark systems. The density of the system has been varied by changing the simulation box size while keeping the total number of atoms fixed at 1000. The entire uniform sample space for both one and two component systems has been illustrated in Fig. 4.

As mentioned earlier in section 2, levels of fidelity in our present study are based on the size of integration timestep of MD simulation. The timestep selected for our present high-fidelity model is 1 fs, which is widely considered a reasonable timestep to produce accurate MD simulation results. On the other hand, for the low fidelity MD model, the timestep is 200 fs, which generally yields very inaccurate MD results. When the size of the timestep is relatively large, the instability of MD simulation integration scheme is one of the major drawbacks. This instability occurs due to high probability of significant overlap between atoms after one timestep integration. Overlap between atoms leads to huge unphysical repulsive forces and large displacements on the next timestep, which in turn results in even more overlap. To prevent this divergence in MD integration scheme for large timestep, we capped the repulsive interactions for atoms very close to each other by modifying the LJ potential at short distance. For all low-fidelity (200 fs timestep) MD simulations, the LJ potential that describes the interatomic interactions has been modified into:

$$u(r_{ij}) = 4\epsilon \left[ \left( \frac{\sigma}{r_{ij}} \right)^{12} - \left( \frac{\sigma}{r_{ij}} \right)^{6} \right] \ for, \frac{\sigma}{r_{ij}} \leq 1.2$$
$$u(r_{ij}) = 0 \ for, \frac{\sigma}{r_{ij}} > 1.2$$

(9)

This capping method has been previously implemented by Razi *et al.*[13] for MD simulation involving large timesteps. For more complex systems, the capping should include modifications in charge-charge, induced dipole- induced dipole and charge-induced dipole interactions. It is worth mentioning that capping may not be required for low temperature systems or cases where low fidelity MD simulation timestep is not very large.

To calculate the atom propagation, the equations of motion are integrated using a velocity verlet algorithm.[36] All the MD simulations of the present study have been performed using LAMMPS[37] and visualization of the present study has been done using OVITO.[38]

Although we have done equal number of high- and low-fidelity MD simulations for this benchmark study, the proposed MPINN is designed such that, one only needs a small number of high-fidelity samples in order to accurately predict a large sample space. All the errors, mean absolute percentage error (MAPE), shown in the present study are calculated using this expression:

$$MAPE = \frac{1}{N} \sum_{i=1}^{N} \left| \frac{A_i - F_i}{A_i} \right|$$

<div align="right">(10)</div>

where, $A_i$ is the system property value obtained from high-fidelity MD simulations and $F_i$ is the system property value obtained from MPINN or low-fidelity MD simulations. Since, this is a benchmark study, the MPINN has been trained 10 times with different initialization values for all the cases considered in the present study. The MAPE values used in the plots of this study are median MAPE over these 10 MPINN prediction errors.

### 3.2 Results & Discussion of the Benchmark Study

The number of high-fidelity data used to train the MPINN does not necessarily need to be same as that of low fidelity data. The number of low fidelity data for each system property is 121, which is equal to the number of points in the sample space. These 121 low fidelity data have been used to train the MPINN for all cases. Along with these low fidelity data, if we use high-fidelity data from 12 uniformly selected sample points to train the MPINN, then it is characterized as an MPINN trained using 10% high-fidelity data. To further illustrate this uniform sampling, the sample points for 10% and 30% high-fidelity sampling are shown in Fig. 4. We have used a wide range of percentages (10% - 90%) of high-fidelity data to train MPINN in order to investigate its performance in predicting system properties of one and two component systems. Here, the evaluated and predicted system properties are total energy per atom, system pressure and diffusion coefficient. We have focused particularly on these three properties since a number of important mechanical[39], thermal[40] and electrical properties[41] can be extracted from them.

### 3.2.1 MPINN performance in one-component system

Figure 5 illustrates the mean absolute percentage errors (MAPE) of the MPINN in predicting different system properties of one-component system for different percentages of high-fidelity training samples. With the increase of high-fidelity training samples, the MAPE decreases for all system properties. This is because, like any other artificial neural network, our present MPINN can predict more accurately with increasing training data. When the high-fidelity training data is 10%, the MAPE for total energy per atom ($E$) is 1%. It is reduced to 0.6% for 20% high-fidelity data, and remains roughly same (0.5%) for 30% and higher high-fidelity training data. This points out that, 30% high-fidelity data is sufficient to accurately predict $E$ of one-component system using our present MPINN. For system pressure ($P$), 30% high-fidelity data is also enough to give accurate prediction (1.3% MAPE) of one component system. In case of diffusion coefficient ($D$), 10% high-fidelity training data produces 8.3% MAPE and it is decreased to 4% when high-fidelity training data percentage increases to 20%. The value of MAPE roughly remains same (3%) from 30% to 50% high-fidelity data in case of $D$ of one component system. For 60% to 90% high-fidelity training data, MAPE remains about 2.5%. Hence it can be concluded that, 30% high-fidelity training data is good enough to accurately train the MPINN to predict $D$ of one-component system.

From the above discussion, it is clear that, 30% high-fidelity training data can be used to train the MPINN in order to give relatively accurate predictions of one-component system properties. To illustrate this, we have presented total energy per atom, system pressure and diffusion coefficient contours over the entire sample space in Fig. 6, 7 and 8, respectively. Fig. 6(a) and (b) show that, the one-component total energy per atom contour of high-fidelity and low fidelity are very different, especially in the low-density region. The MAPE of low fidelity MD simulation $E$ is 25.2%. However, the MPINN, trained with 30% high-fidelity data, decreases the MAPE to 0.5%, which is reflected in the contour of Fig. 6(c). The total energy per atom contours of high-fidelity MD simulation and MPINN prediction trained with 30% high-fidelity data are almost identical, which indicates high predictive power of our present MPINN. In Fig. 7, the one-component system pressure contours over the entire sample space show that, the low and high-fidelity MD simulation results of $P$ are quite different, especially in the higher density and temperature region. MAPE of low fidelity MD simulation is 55.3%. This large error is reduced to 1.3% by the MPINN with 30% high-fidelity training data. As we can see from Fig. 7(a) and (c), the system pressure contours of high-fidelity MD results and MPINN prediction (30% high-fidelity data) are same. Fig. 8 shows that, high-fidelity and low-fidelity diffusion coefficient contours are completely different with 77.6% MAPE of low fidelity MD results. It is also worth mentioning that, in low density region, the low fidelity MD results of $D$ are highly erroneous. MPINN trained with 30% high-fidelity training data reduces this error to 3.3%. If we compare Fig. 8(a) and (c), it can be concluded that, high-fidelity and MPINN predicted diffusion coefficient contours are in very well agreement with each other.

Our present benchmark study on a representative one-component system shows that, with 30% high-fidelity training data, our proposed MPINN can produce highly accurate predictions of important system properties.

*3.2.2 MPINN performance in two-component system*

In Fig. 9, the MAPE of our proposed MPINN in predicting different system properties of two-component system for different percentages of high-fidelity training samples. When the high-fidelity training data is 10%, the value of MAPE associated with total energy per atom ($E$) is 0.35%. With the increase of high-fidelity training data percentage, the accuracy of $E$ gradually increases. However, this increase of accuracy is not significant enough, since even with 90% high-fidelity training data, the MAPE is reduced to only 0.25%. Hence, 10% high-fidelity training data is sufficient for the MPINN to predict $E$ of a two-component system accurately. In case of two component system pressure ($P$), the value of MAPE decreases with the increase in high-fidelity training data. However, after 30% high-fidelity training data where MAPE is 1.3%, adding more training data does not result in any significant increase in accuracy. The MAPE associated with self-diffusion coefficients of components A ($D_A$) and B ($D_B$) varies with high-fidelity training data percentage in a similar manner. When high-fidelity training data is 30%, the MAPE of $D_A$ and $D_B$ are 3.75% and 3.79%, respectively. These values of MAPE decrease to 3.3% at 90% high-fidelity training data. Hence for all four properties of two component system, training with 30% high-fidelity data seems to provide reasonably accurate MPINN predictions.

We have presented the contours of total energy per atom, system pressure and self-diffusion coefficients of two-component system over the entire sample space in Fig. 10-13. Fig. 10 shows that, the MAPE value of low-fidelity $E$ is 4.8%, which is small enough to be considered accurate prediction. From the MAPE value, it may seem like that low-fidelity MD simulations are good enough total energy per atom predictor. However, if we compare the contours of Fig. 10(a) and (b), it is clear that they give very different trends. In high-fidelity MD results, $E$ decreases with increasing density over the entire temperature region. However, low fidelity MD results show that, in high temperature region, the E increases with the increase in density. On the other hand, the MPINN trained with 30% high-fidelity data gives more accurate predictions not only in terms of MAPE (0.3%), but also in terms of trend (Fig. 10(c)). Fig. 11 shows the system pressure contours obtained from high- and low-fidelity MD simulations and MPINN (30% high-fidelity training data). The P values obtained from low fidelity MD simulation have MAPE of 3.6%. It also follows similar trend compared to high-fidelity MD system pressure (Fig. 11(a)). MPINN trained with 30% high-fidelity data reduces the MAPE to 1.3%, and also gives the correct trend, as shown in Fig 11(c). Figure 12(a) and (b) show that, self-diffusion coefficient of component-A ($D_A$) follows the same trend in both high- and low-fidelity MD simulations. However, these figures also point out that, the values are vastly different, with low fidelity MD results having MAPE value of 61.2%. As we can see from Fig. 12(a) and (c), $D_A$ contours of high- fidelity MD results and MPINN prediction (30% high-fidelity data) are roughly same with MAPE value of only 3.8%. Figure 13 illustrates the contours of self-diffusion coefficient of component-B ($D_B$). Low fidelity MD results of $D_B$ have an MAPE value of 25.3%, whereas MPINN trained with 30% high-fidelity data yields only 3.8% MAPE. Also Fig. 13(a) and (c) show that, high-fidelity MD results and MPINN predictions associated with $D_B$ are in well agreement with each other.

Our present benchmark study on a representative two-component system shows that, with 30% high-fidelity training data, our proposed MPINN can produce highly accurate predictions of important system properties. Furthermore, MPINN is capable of learning the trend of system properties with density and temperature with very few high-fidelity training data.

### 3.2.3 Computational Saving

One of the main advantages of our present implementation of MPINN is to achieve high accuracy MD simulation results with significantly less computational cost compared to conventional MD simulation procedure. Figure 14(a) and (b) illustrate the computational saving and its effect on the accuracy of the MPINN predictions of one and two component system properties, respectively. The computational saving (CS) have been calculated using eqn. 11.

$$CS = \frac{1}{N}\sum_{i=1}^{N}\left|\frac{CC_{HF} - CC_{LF}}{CC_{HF}}\right| \qquad (11)$$

where, $CC_{HF}$ and $CC_{LF}$ are the computational cost associated with high- and low-fidelity MD simulations, respectively. We have performed both low and high-fidelity simulations using 1 processor per simulation, in a workstation of 2.40 GHz processor clock speed, 512 GB RAM and 64 bit operating system. The details of these computational costs have been made available in this repository (see Data Availability section). It is clear from Fig.14 that, when the CS is high, the value of MAPE is also high for all cases. For one component system $D$, reducing the CS from 88% to 77% results in 50% decrease in MAPE value. Further decreasing the CS to 68%, results in a 20% drop in MAPE value. This initial high decrease of MAPE value also exists in case of E and P. In the previous sections, we have concluded 30% high-fidelity training data to be an optimum high-fidelity training percentage in case of one-component system. It is seen from Fig. 14(a), that MPINN trained using 30% high-fidelity data results in 68% CS. Similar trend is present in case of two component systems, as shown in Fig. 14(b). When the CS is high, reducing it by adding more high-fidelity data to the MPINN leads to a large decrease in MAPE. But when CS is below 50%, reducing it has no significant effect on the value of MAPE, and thus on the accuracy of the MPINN. For two-component system, MPINN trained using 30% high-fidelity data also results in 68% CS.

These high values of computational saving in both benchmark systems highlight the outstanding feature of our present implementation of deep neural network. Also, our present methodology has potential to enable researchers to explore larger timescales in MD simulations. Another interesting aspect of our present study is that, the computational cost of training the MPINN is very low and remains that way regardless of the computational cost associated with high- and low-fidelity MD simulations. Despite the success demonstrated, a significant limitation of our present methodology is a lack of more sophisticated sampling method in selecting high-fidelity data. Also, energy capping of a more complicated interatomic potential function for low fidelity simulations may add difficulty to the entire methodology.

## 4. Nanofluid viscosity prediction using MPINN

### 4.1 Molecular dynamics methodology

After the benchmark study, we implemented the MPINN to predict the viscosity of argon-copper nanofluid for a wide range of temperature and volume fraction. This implementation of MPINN in predicting nanofluid viscosity has been conducted in order to demonstrate the applicability of our present methodology in actual MD studies. We have particularly chosen argon-copper nanofluid as our system of interest, since its physical properties have been widely studied using MD simulation by the community of nanoparticle research[42–45].

The MD simulation system for calculating nanofluid viscosity consists of argon as base fluid and spherical copper nanoparticle of different sizes as shown in Fig. 15. The size of the simulation box is 68.6 Å × 68.6 Å × 68.6 Å. Initially, argon was arranged in face-centered cubic (FCC) lattice with lattice constant of 5.72 Å, such that the total number of argon atoms is 6914, leading to a density of 1.418 g/$m^3$. Central spherical region of different sizes were carved out by placing FCC copper atoms with lattice constant 3.62 Å. The size of copper nanoparticle is selected based on the desired nanofluid volume fraction (φ). Periodic boundary condition has been applied in all directions in order to mimic volume properties and to avoid boundary effects. The interatomic potential of argon-argon and argon-copper are described by the well-known Lennard–Jones (LJ) 12-6 potential[33] (equation 6) with a cutoff distance of 2.8$\sigma_{Ar}$. The potential parameters (σ and ε) have been presented in table 3.

**Table 3**. LJ (12-6) interatomic potential parameters used for nanofluid viscosity study.

| Particles $(i,j)$ | $\varepsilon_{ij}$(eV) | $\sigma_{ij}$(Å) |
|---|---|---|
| *Ar-Ar* | 0.0104 | 3.405 |
| *Ar-Cu* | 0.0651 | 2.872 |

For describing the interatomic interaction between copper atoms, we have used embedded-atom method (EAM) potential [46] parameterized by Daw and Baskes. In this context, the energy of an atom $i$ is determined by equation 12:

$$E_i = F_i \left( \sum_{i \neq j} \rho(r_{ij}) \right) + \frac{1}{2} \sum_{i \neq j} \varphi_{ij}(r_{ij}) \tag{12}$$

where, $r_{ij}$ is the interatomic distance between atoms $i$ and $j$. $F_i$ is the embedding energy of atom $i$ depending on the atomic electron density, $\rho$ and $\phi$ refer to the short-range pair potential interaction.

Energy minimization using conjugate-gradient (CG) method has been performed for adjusting the initial positions of atoms. Then the system has been equilibrated for 400 ps under canonical ensemble (NVT) with a temperature damping parameter of 100 fs and a drag factor of 0.3. Nose-Hoover thermostat is used to keep temperature constant. After the system has achieved equilibrium state, a relaxation under micro-canonical ensemble (NVE) for 8 ns has been used for data production. The Green-Kubo method has been used to calculate viscosity in the equilibrium state. This approach uses flux fluctuations for viscosity evaluation via the fluctuation-dissipation theorem [47,48]. For high-fidelity simulations the timestep of the MD simulation is 1 fs, whereas it is 20 fs for low fidelity simulations. No energy capping similar to eqn. 9 has been implemented in the low fidelity simulations in order to avoid complications.

We have studied the effects of temperature ($T$) and volume fraction ($\phi$) on the viscosity of argon-copper nanofluid. For this purpose, $T$ from 86 K to 102 K and $\phi$ from 0 to 4% have been taken as the parameter space of this study. We have performed 12 high-fidelity simulations, whereas, 40 low fidelity simulations have been conducted across the whole sample space uniformly. The results obtained from these simulations have been used as the training data of MPINN. The architecture of the MPINN used in this study is same as the one used for the benchmark study (Fig. 2). Since the number of both high- and low-fidelity data are small compared to the benchmark study, the value of regularization parameter used in all three neural networks ($NN_L, NN_{H1}$ and $NN_{H2}$) is 0.02, in order to avoid overfitting, since numbers of both low- and high-fidelity data are relatively small.

### 4.2 Predictive performance of MPINN for nanofluid viscosity

In this section, we will discuss the application of the MPINN in predicting nanofluid viscosity. This is an example of the many possible MD studies that can be accelerated using our proposed methodology. We have conducted 40 low fidelity and 12 high-fidelity MD simulations. The number of high-fidelity simulations is 30% of that of low fidelity simulations. This percentage of high-fidelity data has been chosen since it can provide sufficiently accurate and accelerated predictions as shown in sections 3.2.2 and 3.2.3 above. These MD results are then used to train an MPINN to make predictions across the entire sample space and the results are presented in Fig. 16. It is apparent from Fig. 16(a) that, viscosity increases with increasing nanofluid volume fraction $\phi$, and decreases with the increase of temperature $T$. In both cases, the trend is almost linear. This result is consistent with previous studies [42,44,49] on nanofluids. Therefore, this MPINN, despite having been trained with very few high-fidelity data points, as shown by the red dots on Fig. 16, has successfully predicted the effects of $T$ and $\phi$ on the viscosity of nanofluid.

Relative viscosity is also an important parameter to understand the enhancing effects of nanofluids. It is defined as the ratio of the viscosity of the nanofluid ($\mu_{nf}$) to the base fluid viscosity ($\mu_{bf}$) at the same temperature. Fig. 16(b) shows the variation of relative viscosity with $T$ and $\phi$ across the entire sample space. An interesting observation from the contour is that, the effect of temperature is more prominent for higher volume fractions. On the other hand, at high temperatures, $\phi$ affects the relative viscosity more significantly. Zeroual *et al.* [42] reported similar effects at higher temperatures in their study. It is observed from Fig. 16(b) that, only the 12 high-fidelity data points are clearly not sufficient to understand these effects of $T$ and $\phi$ on relative viscosity, particularly at high $T$ and $\phi$ region. But by using these 12 high-fidelity results as training data along with 40 low fidelity results, the MPINN can accurately predict the effects of $T$ and $\phi$ on relative viscosity. To further confirm the accuracy of our present neural network, we have compared the relative viscosity at 86K with three viscosity models. The details of these models are presented in Table.4.

**Table 4.** Theoretical models of predicting viscosity of nanofluids.

| Name | Equation |
|---|---|
| Einstein model [50] | $\mu_{nf} = (1 + 2.5\,\phi)\mu_{bf}$ |
| Brinkman model [51] | $\mu_{nf} = \frac{1}{(1-\phi)^{2.5}}\,\mu_{bf}$ |
| Batchelor model [52] | $\mu_{nf} = (1 + 2.5\,\phi + 6.2\,\phi^2\mu_{bf})$ |

Figure 17 presents a comparison between the MPINN predicted relative viscosities and the aforementioned theoretical models. It is apparent that, our MPINN predicted MD results are in well agreement with these theoretical models. Therefore, we can conclude that, the MPINN can accurately predict the trends and values of nanofluid viscosity with volume fraction and temperature. Also, Fig. 17 illustrates the difference between low-fidelity MD results and MPINN predictions. Again, this demonstrates the effectiveness of the MPINN to extract important features of nanofluid viscosity from highly erroneous low-fidelity data.

It is worth mentioning that computationally, the low fidelity simulations are approximately 16 times faster than their high-fidelity counterparts. Hence, 40 low-fidelity and 12 high-fidelity simulations are equivalent to roughly 15 high-fidelity simulations, in terms of computational cost. Without MPINN, 15 high-fidelity simulations will not be sufficient to generate the contours of Fig. 16 accurately. Our present methodology is both computationally efficient and accurate in predicting practical MD simulation results. This methodology of predicting viscosity of nanofluids using MPINN can be extended to other nanofluid properties such as thermal conductivity, heat capacity, expansion coefficient etc. with similar computational efficiency. Furthermore, we would like to reiterate that, this is just an example of what can be achieved with this method of using an MPINN, to yield much more detailed MD results across an entire sample space, which has long been deemed nearly impossible due to the limited computational resources.

## 5. Conclusion

In our present study, we have proposed an implementation of multi-fidelity physics informed neural network (MPINN) in large scale MD simulations to reduce computational cost, where the simulation fidelity is based on length of the timestep. Then, the capability of our implemented MPINN has been demonstrated on two benchmark problems involving one and two component Lennard-Jones systems. It is found that, with 30% high-fidelity training data, the MPINN can accurately predict various system properties for a wide range of temperatures and densities. The MPINN can also predict the trends of the system properties with density and temperature very accurately. We also observed that, the high-fidelity and low fidelity results are not required to be close to achieve higher precision with the MPINN. These highly accurate predictions are accompanied by 68% computational saving. The computational cost of training MPINN is very small and does not depend on the complexity of high-fidelity simulations. Our present methodology is the first implementation of MPINN for predicting nanoscale properties using MD simulation, which can pave way to investigate more complex engineering problems that demand long range MD simulations. As a proof of concept, we studied the effects of temperature and volume fraction on viscosity of argon-copper nanofluid using MPINN and then compared them with previous studies and models. It is found that MPINN, with much less computational time, can accurately predict nanofluid viscosity at a wide range of sample space. This demonstrates what can potentially be achieved with this method to determine accurate MD results over large sample spaces with limited computational resources. Furthermore, systems involving complex force fields can potentially be integrated into our present multi-fidelity framework by implementing appropriate energy capping methods which may include modifications in charge-charge, induced dipole-induced dipole and charge-induced dipole interactions.

**Declaration of competing interest**

The authors declare that they have no known competing financial interests or personal relationships that could have appeared to influence the work reported in this paper.

**Acknowledgement**

This work has been carried out in the department of Mechanical Engineering, Bangladesh University of Engineering and Technology (BUET), Dhaka-1000, Bangladesh. The authors gratefully acknowledge the support and facilities provided by BUET.

**Data Availability**

MATLAB scripts and workspaces of the MPINN, along with training data for both benchmark and nanofluid viscosity study are available in this repository. Also, the computational costs associated with all the individual benchmark MD simulations are available in this repository.

| | **List of Figure Captions** |
|---|---|
| **Figure 1** | Time evolution of an LJ system energy per atom for different timesteps, and their corresponding computation times (in brackets). |
| **Figure 2** | Multi-fidelity physics informed neural network (MPINN) used in the present study. Here, $L$ represents low-fidelity and $H$ represents high-fidelity. $NN_{H1}$ is linear neural network, and $NN_{H2}$ is non-linear neural network. $\alpha$ is a hyper-parameter which determines the degree of nonlinearity between low and high-fidelity data. |
| **Figure 3** | Snapshots of **(a)** single-component and **(b)** two-component Lennard Jones systems. |
| **Figure 4** | Entire uniform sample space of density and temperature for both one and two component benchmark study along with 10% and 30% high-fidelity sample points. |
| **Figure 5** | MAPE of MPINN in predicting one-component system properties for different percentages of high fidelity training samples. ($E$ = total energy per atom, $P$ = system pressure, $D$ = self-diffusion coefficient) |
| **Figure 6** | Total system energy per atom (eV/atom) contours of one-component system, over the entire sample space obtained from (a) high fidelity MD simulation, (b) low fidelity MD simulation (MAPE = 25.2%) and (c) MPINN trained using 30% high fidelity data (MAPE = 0.5%). |
| **Figure 7** | System pressure (bar) contours of one-component system, over the entire sample space obtained from (a) high fidelity MD simulation, (b) low fidelity MD simulation (MAPE = 55.3%) and (c) MPINN trained using 30% high fidelity data (MAPE = 1.3%). |
| **Figure 8** | Diffusion coefficient ($Å^2$/ps) contours of one component system, over the entire sample space obtained from **(a)** high fidelity MD simulation, **(b)** low fidelity MD simulation (MAPE: 77.6%) and **(c)** MPINN trained using 30% high fidelity data (MAPE = 3.3%). |
| **Figure 9** | MAPE of MPINN in predicting two-component system properties for different percentages of high fidelity training samples ($E$ = total energy per atom, $P$ = system pressure, $D_A$ = self-diffusion coefficient of component A, $D_B$ = self-diffusion coefficient of component B) |
| **Figure 10** | Total system energy per atom (eV/atom) contours of two-component system, over the entire sample space obtained from **(a)** high fidelity MD simulation, **(b)** low fidelity MD simulation (MAPE: 4.8%) and **(c)** MPINN trained using 30% high fidelity data (MAPE: 0.3%). |
| **Figure 11** | System pressure (bar) contours of two component system, over the entire sample space obtained from **(a)** high fidelity MD simulation, **(b)** low fidelity MD simulation (MAPE: 3.6%) and **(c)** MPINN trained using 30% high fidelity data (MAPE: 1.3%). |
| **Figure 12** | Self-diffusion coefficient of component A ($Å^2$/ps) contours of two-component system, over the entire sample space obtained from **(a)** high fidelity MD simulation, **(b)** low fidelity MD simulation (MAPE = 61.2%) and **(c)** MPINN trained using 30% high fidelity data (MAPE = 3.8%). |
| **Figure 13** | Self-diffusion coefficient of component B ($Å^2$/ps) contours of two-component system, over the entire sample spacee obtained from **(a)** high fidelity MD simulation, **(b)** low fidelity MD simulation (MAPE = 25.3%) and **(c)** MPINN trained using 30% high fidelity data (MAPE = 3.8%). |

| Figure 14 | Computational saving using MPINN and corresponding MAPE in predicting **(a)** one-component and **(b)** two-component system properties ($E$ = total energy per atom, $P$ = system pressure, $D$ = self-diffusion coefficient). |
|---|---|
| Figure 15 | Snapshot of the initial configuration of *Ar-Cu* nanofluid simulation box for viscosity measurements. |
| Figure 16 | **(a)** Viscosity and **(b)** Relative Viscosity of *Ar-Cu* nanofluid for different volume fractions at various temperatures, predicted by MPINN trained with few high-fidelity data (red dots). |
| Figure 17 | Relative Viscosity of *Ar-Cu* nanofluid for different volume fractions at 86K, predicted by MPINN, compared to previous models and different fidelity MD results. |

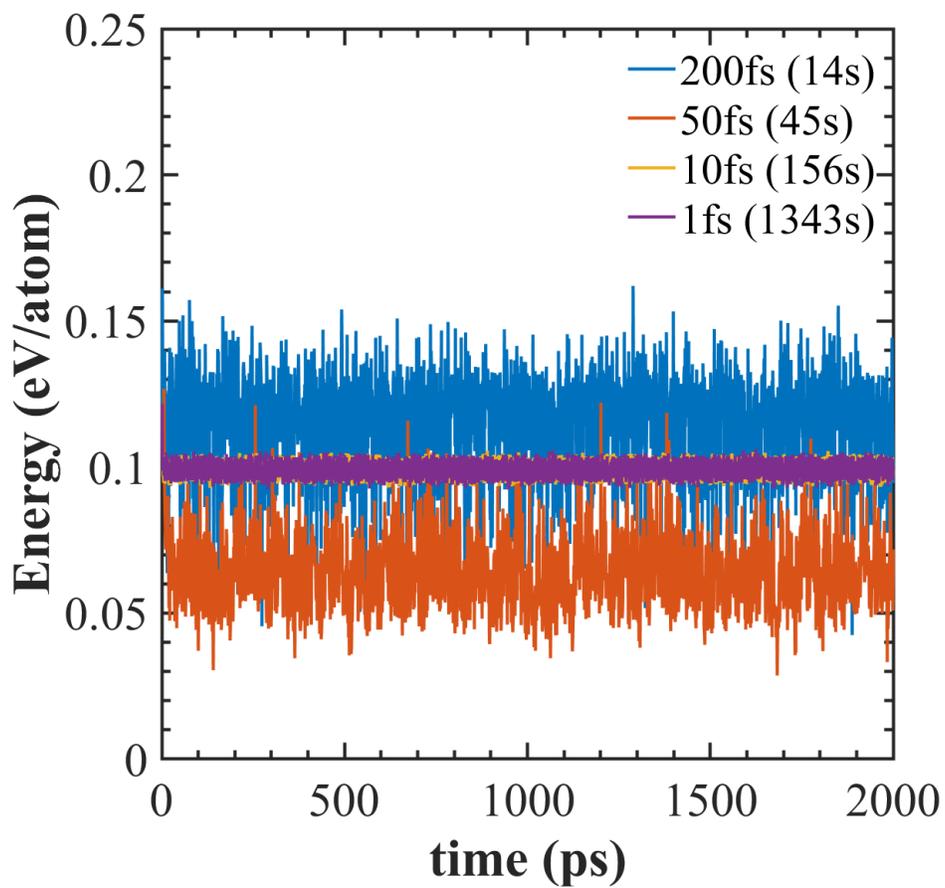

**Figure 1.** Time evolution of an LJ system energy per atom for different timesteps, and their corresponding computation times (in brackets).

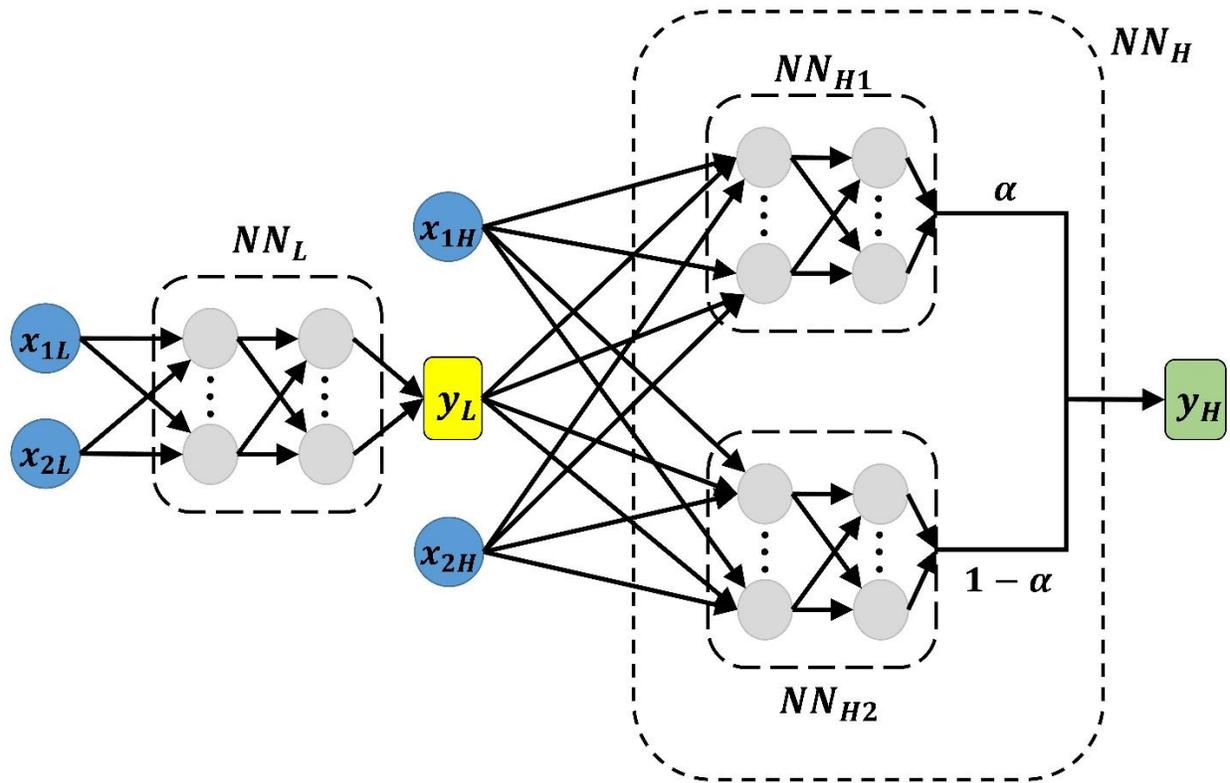

**Figure 2.** Multi-fidelity physics informed neural network (MPINN) used in the present study. Here, *L* represents low-fidelity and *H* represents high-fidelity. *NN_H1* is linear neural network, and *NN_H2* is non-linear neural network. $\alpha$ is a hyper-parameter which determines the degree of nonlinearity between low and high-fidelity data.

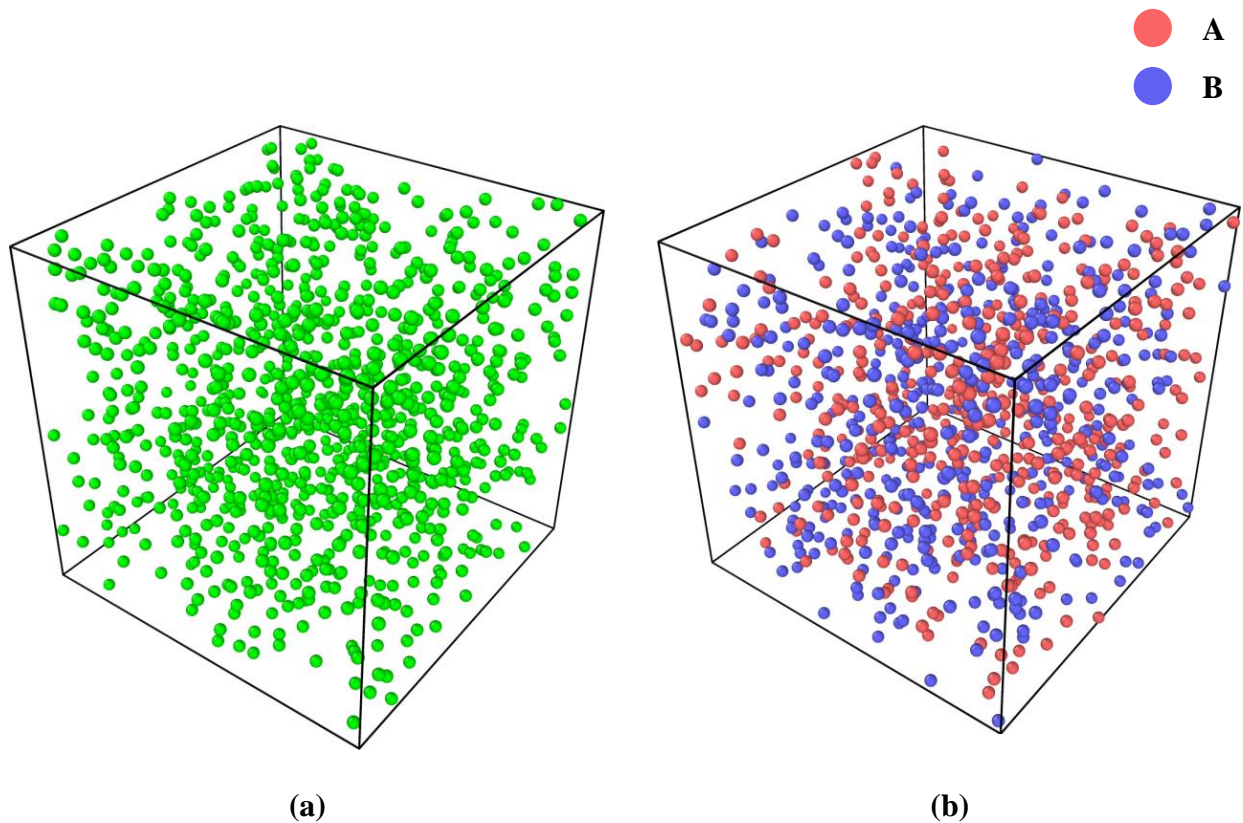

**(a)**                                **(b)**

**Figure 3.** Snapshots of **(a)** single-component and **(b)** two-component Lennard Jones systems.

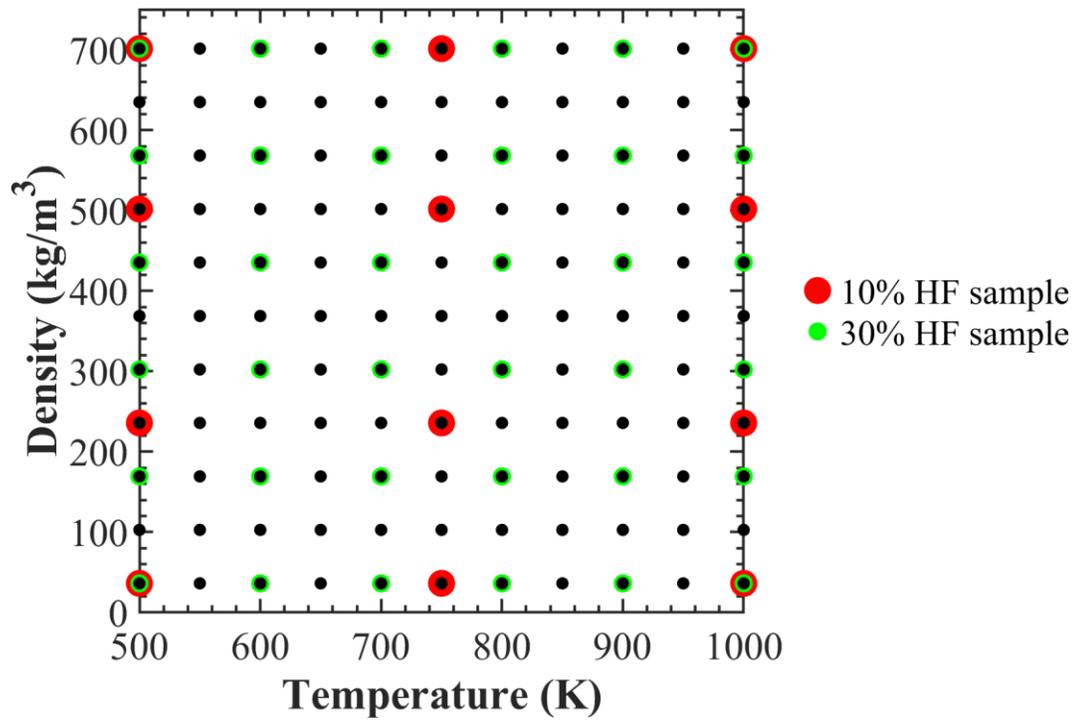

**Figure 4.** Entire uniform sample space of density and temperature for both one and two component benchmark study along with 10% and 30% high-fidelity sample points.

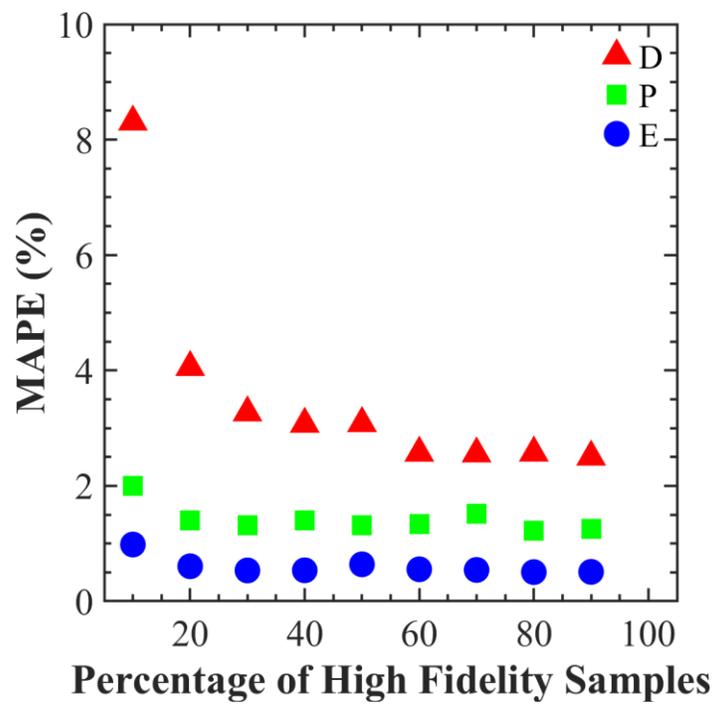

**Figure 5.** MAPE of MPINN in predicting one-component system properties for different percentages of high fidelity training samples. (*E* = system energy per atom, *P* = system pressure, *D* = self-diffusion coefficient)

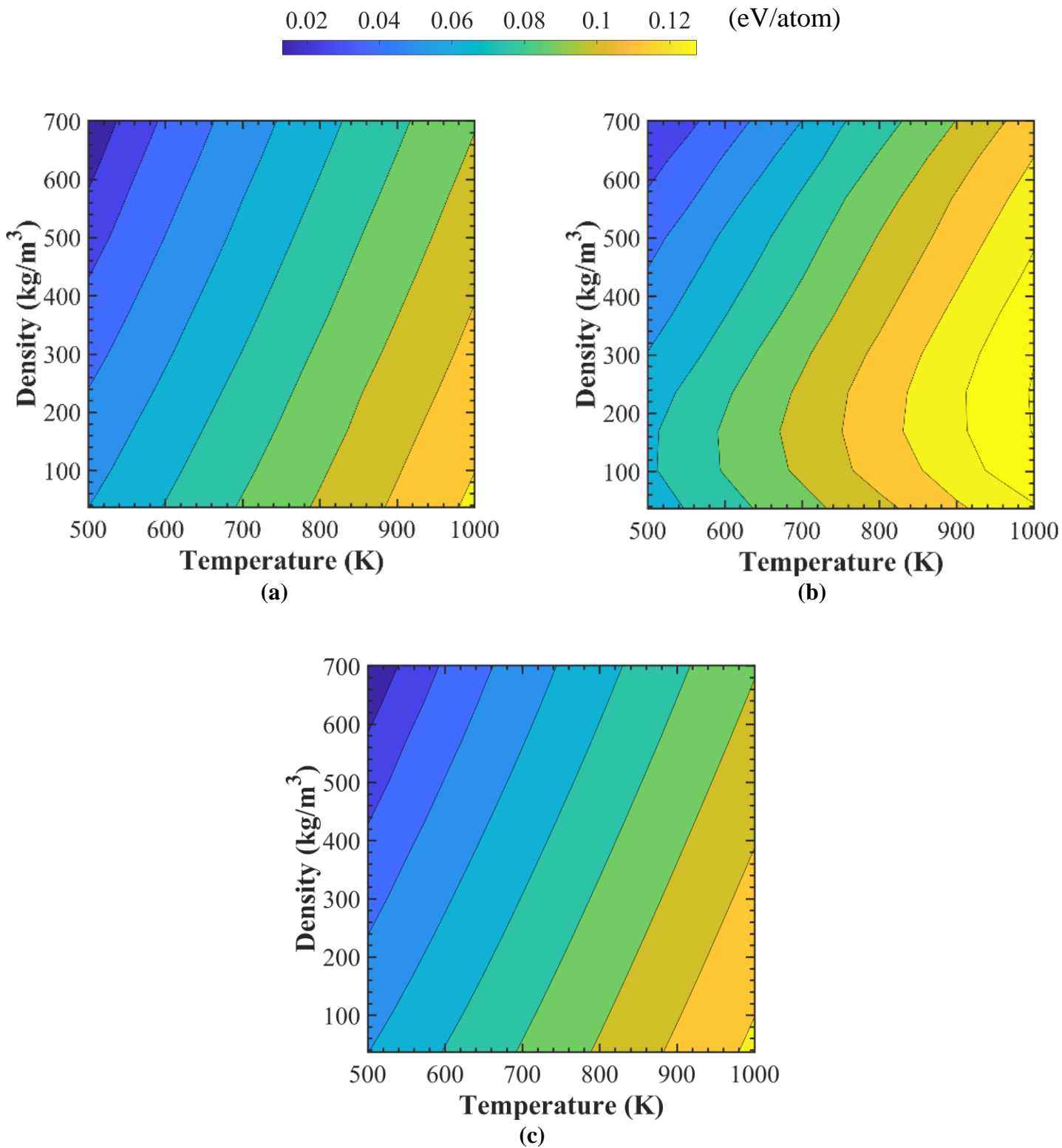

**Figure 6.** Total system energy per atom (eV/atom) contours of one-component system, over the entire sample space obtained from **(a)** high fidelity MD simulation, **(b)** low fidelity MD simulation (MAPE = 25.2%) and **(c)** MPINN trained using 30% high fidelity data (MAPE = 0.5%).

**Figure 7.** System pressure (bar) contours of one-component system, over the entire sample space obtained from **(a)** high fidelity MD simulation, **(b)** low fidelity MD simulation (MAPE = 55.3%) and **(c)** MPINN trained using 30% high fidelity data (MAPE = 1.3%).

**Figure 8.** Diffusion coefficient ($Å^2$/ps) contours of one component system, over the entire sample space obtained from **(a)** high fidelity MD simulation, **(b)** low fidelity MD simulation (MAPE: 77.6%) and **(c)** MPINN trained using 30% high fidelity data (MAPE = 3.3%).

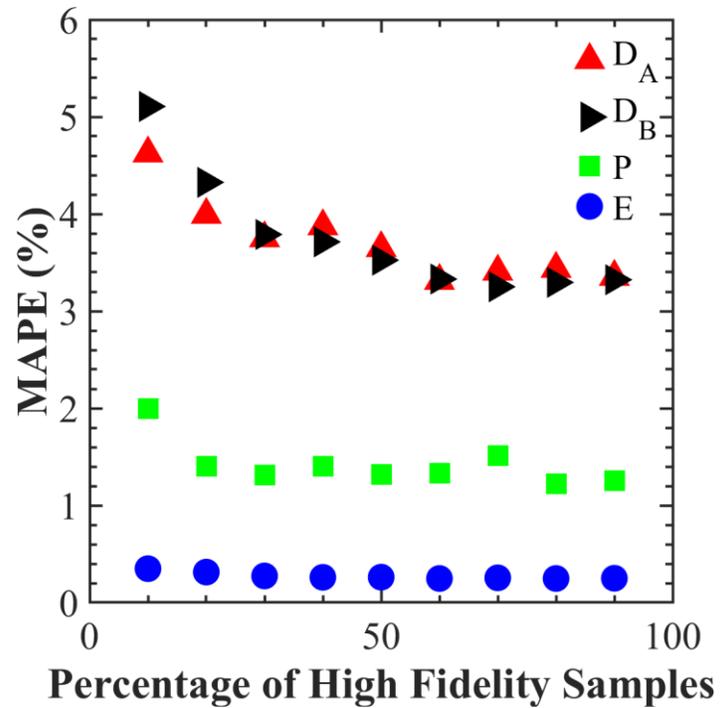

**Figure 9.** MAPE of MPINN in predicting two-component system properties for different percentages of high fidelity training samples (E = total energy per atom, P = system pressure, $D_A$ = self-diffusion coefficient of component A, $D_B$ = self-diffusion coefficient of component B)

**Figure 10.** Total system energy per atom (eV/atom) contours of two-component system, over the entire sample space obtained from **(a)** high fidelity MD simulation, **(b)** low fidelity MD simulation (MAPE: 4.8%) and **(c)** MPINN trained using 30% high fidelity data (MAPE: 0.3%).

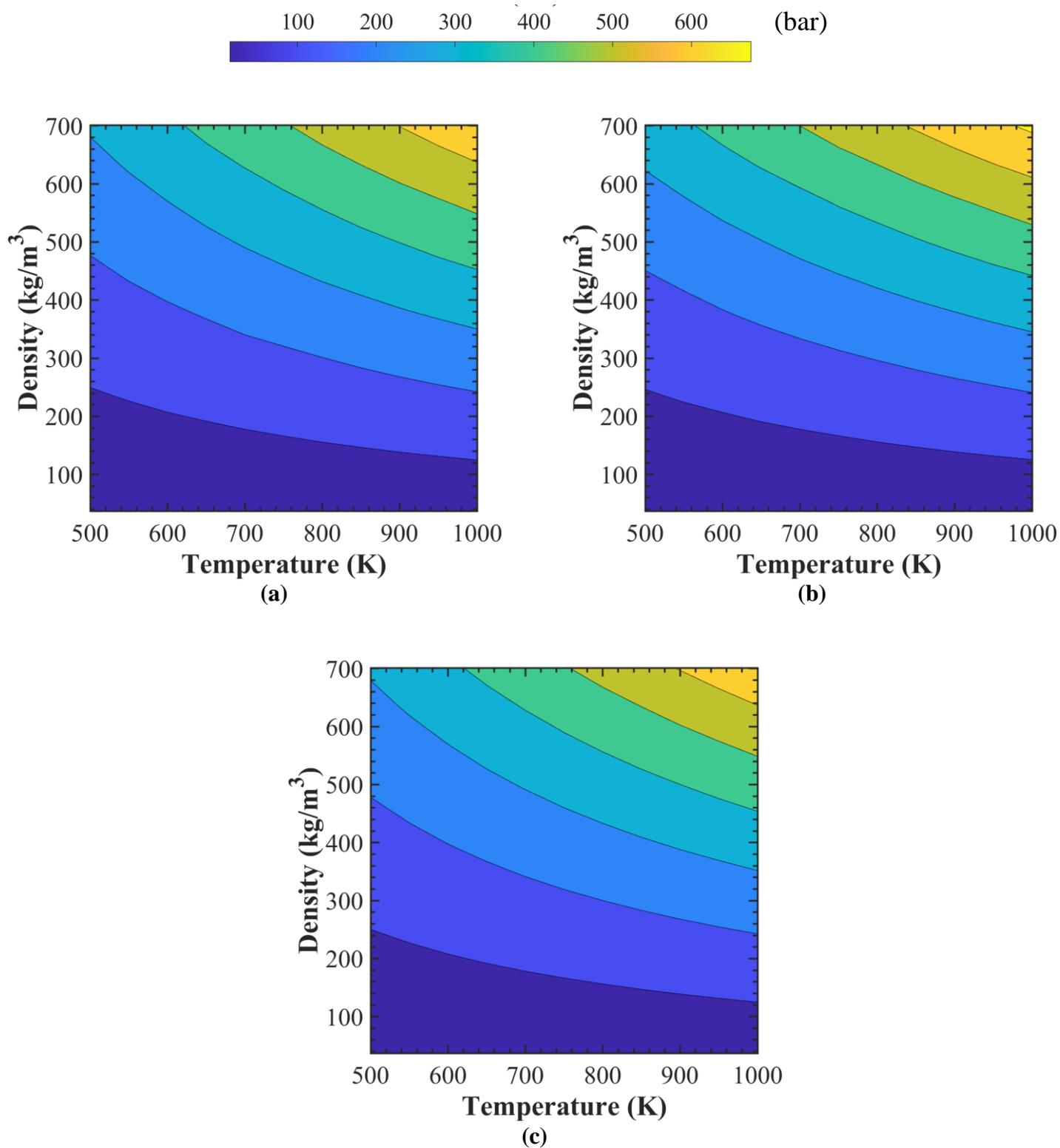

**Figure 11.** System pressure (bar) contours of two component system, over the entire sample space obtained from (**a**) high fidelity MD simulation, (**b**) low fidelity MD simulation (MAPE: 3.6%) and (**c**) MPINN trained using 30% high fidelity data (MAPE: 1.3%).

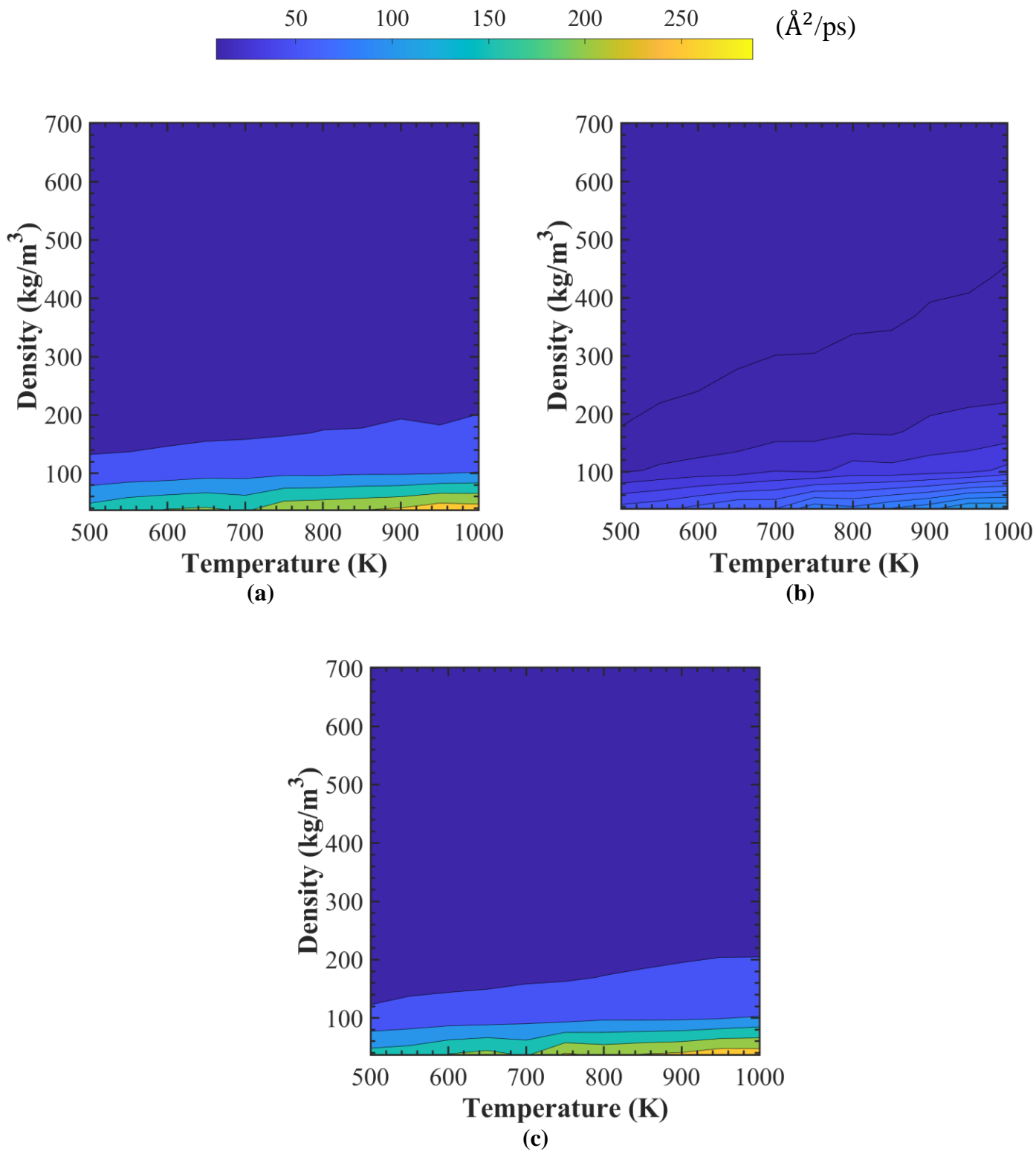

**Figure 12.** Self-diffusion coefficient of component A (Å²/ps) contours of two-component system, over the entire sample space obtained from **(a)** high fidelity MD simulation, **(b)** low fidelity MD simulation (MAPE = 61.2%) and **(c)** MPINN trained using 30% high fidelity data (MAPE = 3.8%).

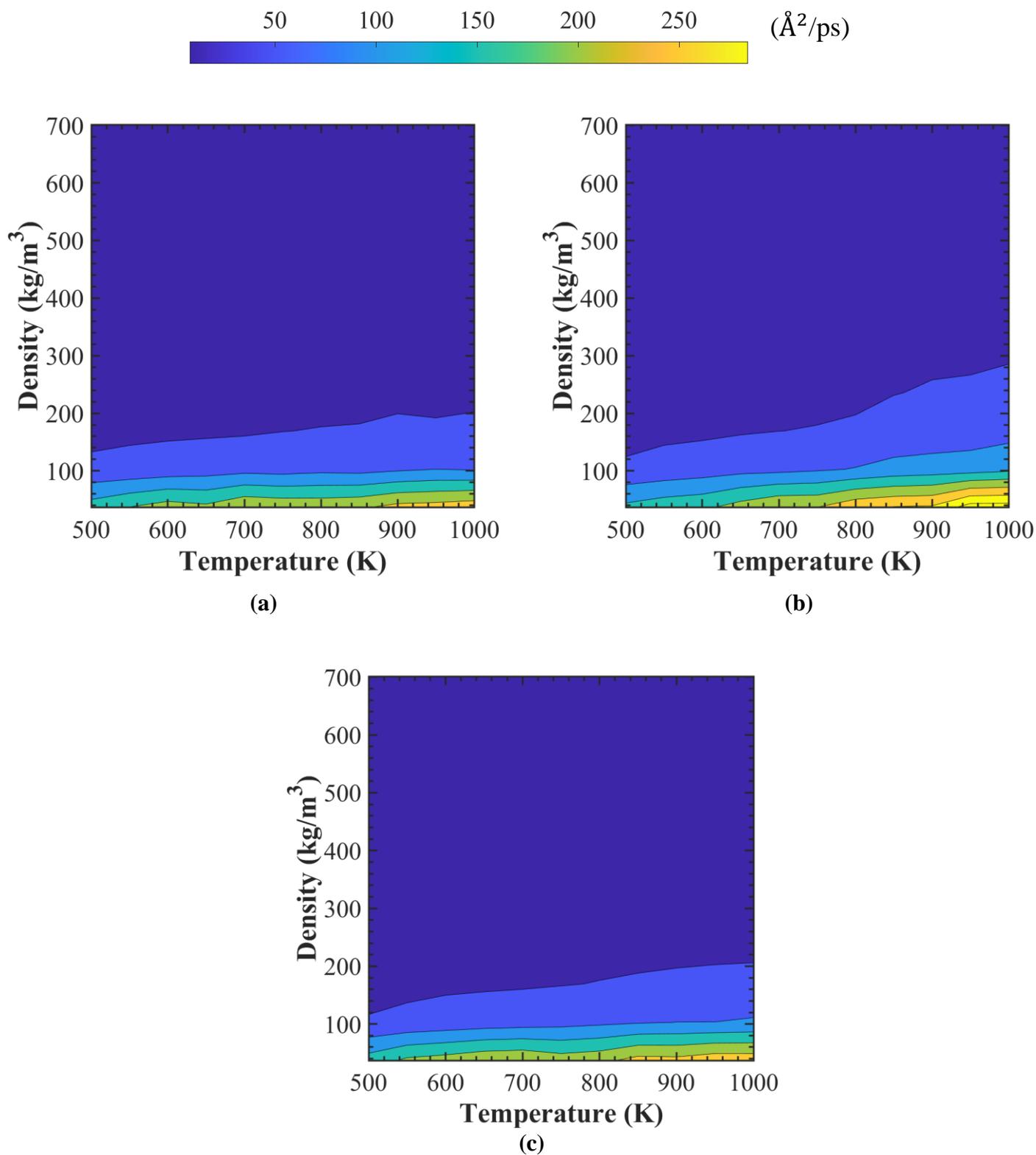

**Figure 13.** Self-diffusion coefficient of component B ($Å^2$/ps) contours of two-component system, over the entire sample spacee obtained from **(a)** high fidelity MD simulation, **(b)** low fidelity MD simulation (MAPE = 25.3%) and **(c)** MPINN trained using 30% high fidelity data (MAPE = 3.8%).

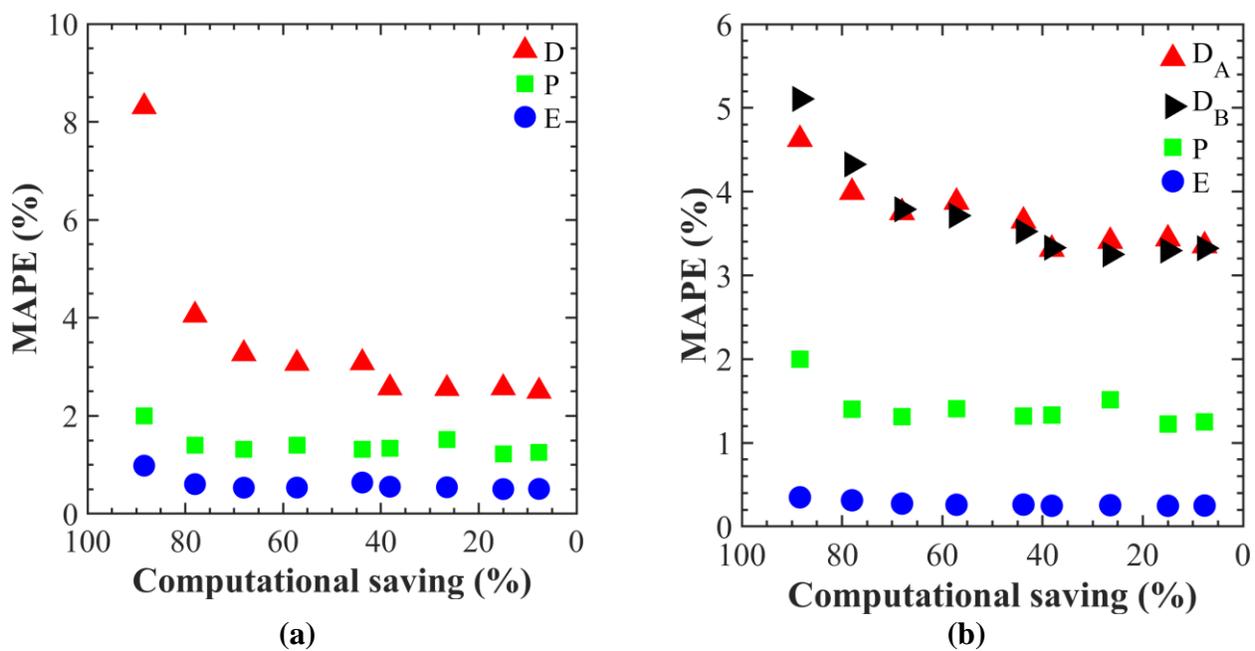

**Figure 14.** Computational saving using MPINN and corresponding MAPE in predicting **(a)** one-component and **(b)** two-component system properties (E = system energy per atom, P = system pressure, D = self-diffusion coefficient).

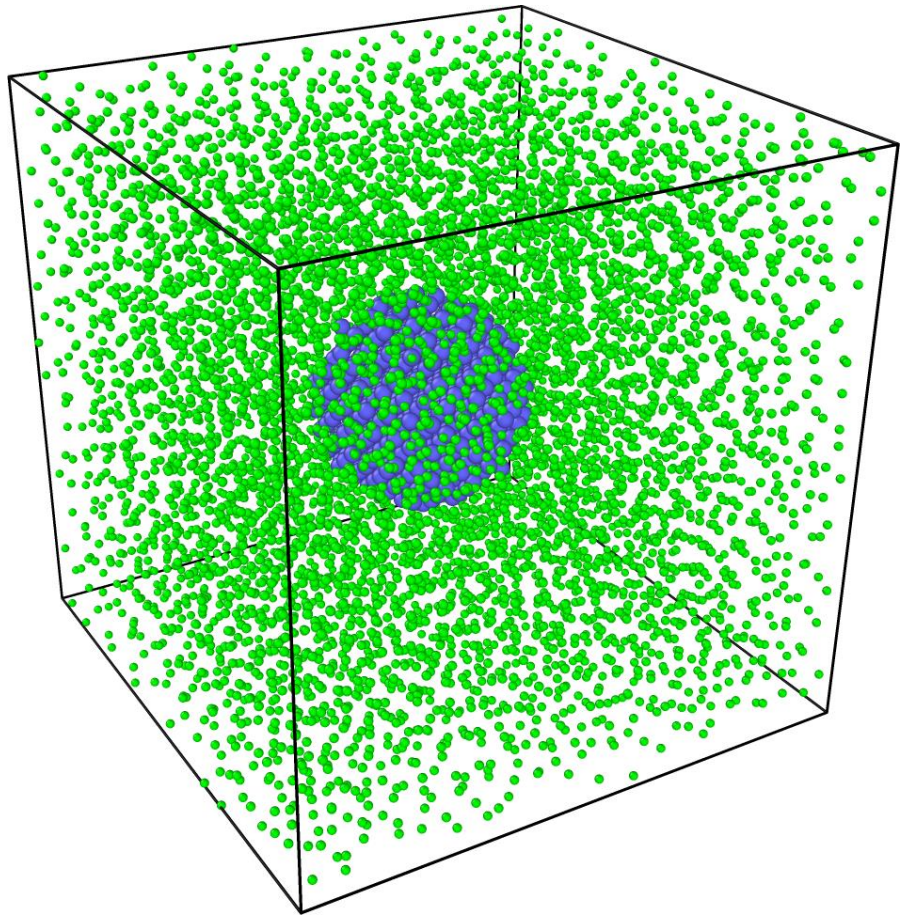

**Fig. 15** Snapshot of the initial configuration of *Ar-Cu* nanofluid simulation box for viscosity measurements.

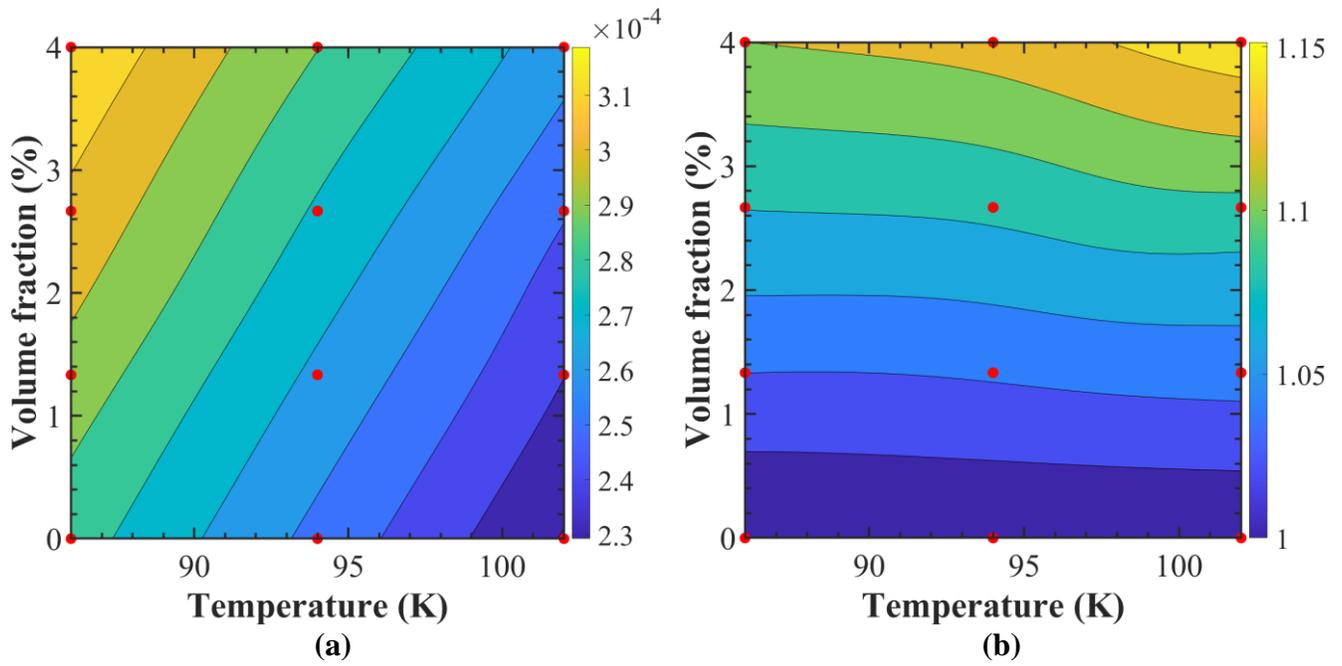

**Figure 16.** (**a**) Viscosity and (**b**) Relative Viscosity of *Ar-Cu* nanofluid for different volume fractions at various temperatures, predicted by MPINN trained with few high-fidelity data (red dots).

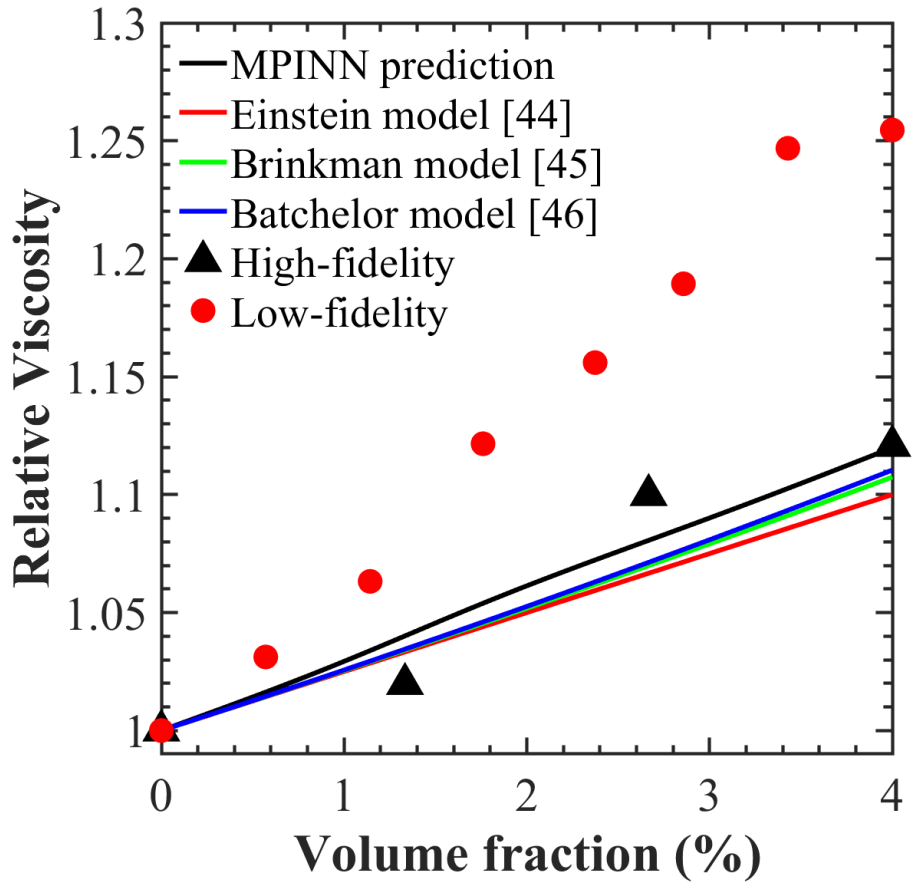

**Figure 17.** Relative Viscosity of *Ar-Cu* nanofluid for different volume fractions at 86K, predicted by MPINN, compared to previous models and different fidelity MD results.

# Author Contribution Statement

Mahmudul Islam: Conceptualization, Methodology, Software, Data Curation, Formal analysis, Writing – Original Draft. Md Shajedul Hoque Thakur: Software, Data Curation, Visualization, Writing-Review & Editing. Satyajit Mojumder: Formal analysis, Writing-Review & Editing. Mohammad Nasim Hasan: Project administration, Resources, Writing - Review & Editing.